\documentclass{article}
\usepackage{parskip}
\usepackage{setspace}
\usepackage{amsmath}
\usepackage{amssymb}
\usepackage{nccmath}
\usepackage{geometry}
\usepackage{hyperref}
\usepackage{authblk}
\usepackage{graphicx}
\usepackage{float}
\usepackage{comment}
\usepackage{xcolor}
\definecolor{darkgreen}{rgb}{0.0, 0.6, 0.5}

\usepackage[
backend=biber,
style=numeric,
sorting=none
]{biblatex}
\title{SPARC: Staking Performance And Reward Coopetition}

\author{Michael D. Norman, Ph.D.\thanks{\href{mailto:me@mdnorman.com}{\texttt{me@mdnorman.com}}} }
\author{Simon Brown}
\author{Mallesh Pai, Ph.D.}
\author{Laurence Smith}
\affil{Consensys}
\date{May 2025}

\begin{document}

\maketitle

\begin{abstract}
This paper presents a novel staking coopetition design aimed at incentivizing decentralization and continuous growth of economic security within a proof-of-stake system. Staking rewards follow a nonlinear mapping relative to stake size. This affords the highest effective yields to smaller operators, fueling network growth and giving users an incentive to delegate their stake to smaller operators. This prevents the preferential accrual and centralization of stake seen in popular blockchains such as Ethereum, where popular liquid staking protocols control large fractions of the total stake thereby having outsized potential impacts on the economic security of the protocol. The proposed system addresses key challenges such as Sybil attacks and offers a comprehensive framework for future research and implementation. We introduce innovative mechanisms and gamification elements, to enhance user engagement and provide transparency in emissions.
\end{abstract}

\section{Introduction}
The evolution of proof-of-stake systems necessitates innovative approaches to incentivize decentralization and enhance economic security. This paper introduces the SPARC - Staking Performance and Reward Coopetition - mechanism, which aims to address these objectives through a competitive-cooperative staking environment \cite{norman2018emergence}. By leveraging stake delegation and gamification, SPARC fosters user engagement and transparency, while mitigating common challenges such as Sybil attacks.

In traditional proof-of-stake (PoS) systems, economic security is often concentrated among a few large stakeholders, leading to centralization risks. There have been previous studies of PoS cryptocurrency networks which have highlighted a recurring trend of accelerating preferential accrual as a result of how rewards are calculated \cite{10328590} \cite{grandjean2024ethereum}, and there have been some proposals for alternatives to mitigate such centralization risks \cite{10634400}.

Overall, existing mechanisms lack sufficient incentives for decentralization and continuous growth. The introduction of a novel staking coopetition design seeks to address these limitations by promoting an optimal (i.e., flatter) stake distribution and continuously enhancing cryptoeconomic security.

\section{Overview}

In SPARC, stakers must lock a minimum amount of the network's native token to participate.  The SPARC mechanism can be applied to any functionally decentralized system, but for ease of discussion, we will refer to the SPARC infrastructure participants as validators and assume their goal is to come to consensus on proposed block validity. 

For each consensus slot, a fixed-size committee of validators is selected at random from the entire set of eligible stakers. These validators are then ordered by the size of their stake and assigned to fixed protocol-defined tiers.

Each tier receives a predefined share of the total slot reward, which is distributed equally among that tier's members. This tier-based structure ensures that higher-staked validators are more likely to be assigned to higher-reward tiers, while enabling lower-staked participants to earn meaningful rewards.

As the selection of validators for each slot is irrespective of the amount they have staked, a rational delegator will delegate to a validator with a lower amount of stake as they will have a higher proportionate share of any rewards that validator earns, while having equal chance of being selected.  This encourages delegators to delegate to validators with lower stake, which results in a flatter distribution of stake over time, thus increasing the protocol's economic decentralization, and in turn, its economic security.

The protocol is deterministic and transparent: the number of tiers, the number of participants per tier, and the reward allocations are constant and specified at the protocol level. This structure enables consistent and analyzable economic incentives across time.

SPARC offers a novel middle ground between egalitarian reward schemes and stake-proportional mechanisms. By blending randomness, stake-based ordering, and tiered incentives, SPARC encourages greater stake commitment while preserving fairness and decentralization. The deterministic tiering and reward logic also facilitate straightforward formal analysis and on-chain enforcement.

\subsection{Mechanism Overview}

The SPARC mechanism can be described at a high-level in relatively simple terms:

\begin{enumerate}
    \item Validators stake their tokens to a minimum amount in order to be eligible to participate in the protocol. Users can delegate their stake to validators.
    \item For each slot, the protocol randomly selects a fixed number of stakers from the global set of validators.
    \item The selected stakers are then ordered by the amount they have staked.
    \item This ordered set of stakers selected for a slot are then grouped into predefined tiers.
    \item Rewards are distributed to stakers based on fixed per-tier allocations, with equal division among members in each tier.
\end{enumerate}

\subsection{Formal Description}

The following section formally describes the mechanism using the aforementioned 5 step sequence.  This results in a simple algorithm that can be applied to any PoS consensus protocol.

\subsubsection*{Variables and Parameters:}

\vspace{10pt}

\begin{tabular}{@{}ll@{}}
\hspace{1em}\textbf{$\mathcal{S}$}      & total set of stakers in the network. \\
\hspace{1em}\textbf{$s_i$}              & individual staker in $\mathcal{S}$. \\
\hspace{1em}\textbf{$T(s_i)$}           & amount of tokens staked by $s_i$. \\
\hspace{1em}\textbf{$S_{\text{min}}$}   & minimum stake required to participate in the protocol. \\
\hspace{1em}\textbf{$x$}                & number of stakers randomly selected per slot. \\
\hspace{1em}\textbf{$k$}                & number of tiers defined in the protocol. \\
\hspace{1em}\textbf{$m_j$}              & number of stakers in tier $j$, where $1 \leq j \leq k$. \\
\hspace{1em}\textbf{$R_j$}              & total reward allocated to tier $j$, where $1 \leq j \leq k$. \\
\hspace{1em}\textbf{$r_{ij}$}           & reward for staker $i$ in tier $j$. \\
\hspace{1em}\textbf{$f(g)$}             & random selection function to choose $x$ stakers from $\mathcal{S}$. \\
\hspace{1em}\textbf{$\mathcal{O}$}      & ordered set of selected stakers sorted by $T(s_i)$, in descending order. \\
\end{tabular}

\subsubsection*{Algorithm Description:}

\begin{enumerate}
    \item Each staker $s_i \in \mathcal{S}$ is eligible if they have minimum amount of stake: $T(s_i) \geq S_{\text{min}}$
    \begin{spacing}{0}
    The global set of eligible stakers thus defined as: $
    \mathcal{S}_{\text{eligible}} = \{ s_i \in \mathcal{S} \mid T(s_i) \geq S_{\text{min}} \}$
    \end{spacing}

    \item The protocol randomly select $x$ stakers for the slot: $\mathcal{S}_{\text{selected}} = f(\mathcal{S}_{\text{eligible}}, x)$

    \item Selected stakers are sorted in descending order by their amount of stake: $\mathcal{O} = \text{Sort}(\mathcal{S}_{\text{selected}}, T(s_i))$

    \item The ordered set $\mathcal{O}$ is then subdivided into $k$ tiers according to their stake-rank in the slot, i.e., top $x_1$ stakers among the selected go into tier $1$, next $x_2$ stakers go into tier $2$ etc.:
    \begin{fleqn}
    \[
    Tier_j = \{ s_i \in \mathcal{O} \mid \text{first } m_j \text{ stakers} \} \quad \text{and} \quad \sum_{j=1}^k m_j = x
    \]
    \end{fleqn}
    The union across all tiers in a slot thus equals:
    \begin{fleqn}
    \[
    \mathcal{O} = \bigcup_{j=1}^k Tier_j \quad \text{and} \quad |Tier_j| = m_j
    \]
    \end{fleqn}

    \item Rewards are distributed equally within each tier:
    \begin{fleqn}
    \[
    r_{ij} = \frac{R_j}{m_j}, \quad \forall s_i \in Tier_j
    \]
    \end{fleqn}
    The total rewards distributed to stakers across all tiers in a slot should equal:
    \begin{fleqn}
    \[
    \sum_{j=1}^k R_j = R_{\text{total}}
    \]
    \end{fleqn}
    where $R_{\text{total}}$ is the total reward available for a slot, as defined by protocol parametrization.
\end{enumerate}

A simple result of the tiering system is that rewards are no longer distributed pro-rata based on the absolute staking amounts; instead they are distributed based on the tier definitions and relative differences in stake of the randomly selected subset. The relative difference forms the basis for the competitive aspects of the coopetition. Formally, the expected rewards of staker $i$ under a staking system like Ethereum would be: 
\[ R \frac{s_i}{\sum_{\ell=1}^S s_\ell }.\]
However, under the tiered rewards system, the probability of a staker being selected in a given period only depends on the number of validators:
\begin{align*}
    \mathbb{P}[\text{selected}] = \frac{x}{S}.
\end{align*}
Next, we can calculate the probability of staker $i$, conditional on selection, being put into tier $j$. Formally, this would require all $j-1$ higher-reward tiers to contain stakers with a higher stake, i.e, somewhere between $\sum_{\ell = 1}^{j-1} x_\ell $ and $\sum_{\ell=1}^{j-1} x_\ell + x_j - 1$ stakers with higher stake to be selected in that round. Since stakers are numbered by stake this means that staker $i$ gets put into tier $j$ if and only if $\sum_{\ell = 1}^{j-1} x_\ell $ and $\sum_{\ell=1}^{j-1} x_\ell + x_j - 1$  of stakers numbered $1$ through $i-1$ are selected in that round. 

Standard probabilistic analysis tells us that this probability can be given by:  
\begin{equation} P\left(\sum_{\ell = 1}^{j-1} x_\ell \leq X \leq \sum_{\ell=1}^{j-1} x_\ell + x_j - 1 \right) = \sum_{k=\sum_{\ell=1}^{j-1} x_\ell }^{\left(\sum_{\ell=1}^{j-1} x_\ell\right) + x_j - 1} \frac{\binom{i-1}{k} \binom{S-i}{x-1-k}}{\binom{S-1}{x-1}} \end{equation}
where: 
\begin{itemize} 
\item X is the random variable representing the number of elements selected from the subset $\{1, \ldots, i-1\}$. 
\item S-1 is the total number of elements to choose from 
\item x is the number of elements being selected (plus 1) \item i is the size of the subset (plus 1) 
\end{itemize}

\vspace{10pt}

By combining this probability with the reward assigned to each tier \( R_j \), we can therefore calculate the expected per slot reward for any validator as:

\[
\mathbb{E}[R_i] = \sum_{j=1}^k P(\text{validator } i \text{ in tier } j) \cdot \frac{R_j}{x_j}
\]

where $x_j$ is the number of validators in tier $j$ and $R_j$ is the number of validators is the total reward allocated to tier $j$ for a given slot.

This deterministic reward modeling is critical in order for validators to participate in the protocol, by allowing them to make decisions based on analyzing the marginal reward benefit of increasing their stake (i.e., moving from rank $i$ to $i-1$). This can also be used by protocol designers to detect potential discontinuities that might encourage gaming, such as splitting or merging their stake to cross tier thresholds.

\subsection{Illustrative Example}

Without loss of generalization, assume Sequencer A has 100 tokens staked and the reward per slot is 10 tokens. Tier 1 receives 5 tokens, Tier 2 receives 3 tokens, and Tier 3 receives 2 tokens. Sequencer A is picked in two separate slots along with other sequencers.

\subsection*{Slot 1}
\begin{itemize}
    \item Sequencer A - 100 tokens staked - Tier 3 Reward - 2 tokens
    \item Sequencer B - 400 tokens staked - Tier 2 Reward - 3 tokens
    \item Sequencer C - 500 tokens staked - Tier 1 Reward - 5 tokens
\end{itemize}

\subsection*{Slot 2}
\begin{itemize}
    \item Sequencer A - 100 tokens staked - Tier 1 Reward - 5 tokens
    \item Sequencer D - 50 tokens staked - Tier 2 Reward - 3 tokens
    \item Sequencer E - 10 tokens staked - Tier 3 Reward - 2 tokens
\end{itemize}

\vspace{10pt}

This example illustrates how the return per token staked varies based on the composition of the selected sequencers, promoting decentralization and engagement.

\section{Results of Simulations}

We tested the SPARC model with a number of different configurations and ran repeated simulations in order to compare the results. Our results show that the model is quite sensitive to it's initial configuration parameters, but that outcomes are consistent for each configuration across repeated simulations.

Our methodology involved establishing a benchmark configuration for a standard PoS consensus protocol based on Tendermint \cite{buchman2016tendermint}, and designing 10 separate experiments of different parameters with which to apply the SPARC model to. This benchmarking facilitated a comparison between standard PoS and SPARC from a delegator's perspective, with regard to delegating to a validator with lower stake versus higher stake.  Our hypothesis is that SPARC incentivizes delegators to delegate to stakers with less stake, with the result of creating a fairer and more optimal distribution of rewards over time.

We ran a simulation of the model 10 times for each of the 10 design experiments, and for each simulation we compared the distribution of tokens over the validator set before and after running a defined number of blocks, correlating to a period of 30 days.

The experiments were deemed "successful" based on whether or not the simulation involving the set of parameters resulted in a significantly more even distribution of token rewards than that of a standard PoS protocol.

\vspace{12pt}

All 10 experiments have different configuration for number of tiers and rewards per tier, but all share the same base configuration, defined with the following constants:

\begin{itemize}
    \item Committee Size (i.e., number of validators per slot): 20
    \item Total number of validators in entire validator set: 1,000
    \item Total token supply: 21,000,000
    \item Total amount staked: 60\%
    \item Annualised issuance rate: 5\%
    \item Slot time: 12 seconds
\end{itemize}

Assuming a 12 second slot time, this amounts to 7,200 slots per day. Therefore, the rewards per slot are calculated as:

\begin{itemize}
    \item \texttt{slots\_per\_year} = 2,628,000
    \item \texttt{total\_staked\_tokens} = \texttt{total\_supply} $\times$ \texttt{total\_amount\_staked} = 21M $\times$ 0.6 = 12.6M
    \item \texttt{annualised\_issuance\_amount} = \texttt{total\_staked\_tokens} $\times$ 0.05 (i.e., 5\% APR)
    \item \texttt{total\_block\_reward} = \texttt{annualised\_issuance\_amount} $\div$ \texttt{slots\_per\_year}
\end{itemize}

\vspace{12pt}
    
\textbf{Standard PoS slot reward calculation:}

\begin{itemize}
    \item Validator reward = \(\texttt{total\_block\_reward} \times \texttt{validator\_stake} \div {\texttt{total\_staked\_tokens}}\)
\end{itemize}

Our tests were carried out with the above constants and by using the sets of parameters list in Table \ref{tab:doe_parameters_and_results}.  The result in the right-most column is in relation to whether or not the simulation involving the set of parameters resulted in a significantly more even distribution of token rewards than that of a standard PoS protocol.

\begin{table}[H]
\centering
\begin{tabular}{|r|r|r|r|c|}
\hline
\textbf{Design Point} & \textbf{Tiers} & \textbf{\# Validators per Tier} & \textbf{Rewards (percentages)} & \textbf{Result} \\ \hline
1 & 1 & 20 & 5 & \textcolor{red}{Fail} \\ \hline
2 & 2 & 10, 10 & 6, 4 & \textcolor{red}{Fail} \\ \hline
3 & 2 & 12, 8 & 6, 3.5 & \textcolor{red}{Fail} \\ \hline
4 & 3 & 7, 7, 6 & 9, 4, 1.5 & \textcolor{red}{Fail} \\ \hline
5 & 3 & 13, 6, 1 & 6, 3.5, 1 & \textcolor{darkgreen}{Success} \\ \hline
6 & 3 & 10, 7, 3 & 8.3, 2, 1 & \textcolor{red}{Fail} \\ \hline
7 & 4 & 10, 6, 3, 1 & 7, 4, 1, 3 & \textcolor{darkgreen}{Success} \\ \hline
8 & 4 & 5, 5, 5, 5 & 10, 5, 3, 2 & \textcolor{red}{Fail} \\ \hline
9 & 5 & 10, 4, 3, 2, 1 & 7, 4, 3, 2, 1 & \textcolor{darkgreen}{Success} \\ \hline
10 & 5 & 6, 8, 3, 2, 1 & 7, 5, 4, 2, 2 & \textcolor{darkgreen}{Success} \\ \hline
\end{tabular}
\caption{DOE Parameters and Results}
\label{tab:doe_parameters_and_results}
\end{table}

\section{Simulation Results}

The design points listed table \ref{tab:doe_parameters_and_results} were simulated with the specified parameters and visualized in the charts below. Each design point output three charts:

\begin{itemize}
    \item Line Chart
    \item Box Plot Chart
    \item Stacked Bar Chart
\end{itemize}

A description of how to interpret the simulation results via the representation in the charts is outlined in the following section:

\subsection{Interpretation of Charts}

\subsubsection{Line Chart}

The line chart that plots the initial distribution of tokens across the validator set in blue, and the resulting final distribution after the simulation has completed in green. This chart also marks the highest number of tokens staked in both the initial and final distributions as vertical lines.  The horizontal axis of the line chart represents the amount of tokens held by one or more validators, and the vertical axis represents the number of validators that have that respective amount of tokens held.

A simulation is deemed successful if it results in a flatter distribution of rewards across all stakers, with at least a marginal bias toward favoring stakers with small stakes. An optimal distribution of rewards looks like a tall, thin, narrow spike around a certain point on the x-axis. This represents a large number of stakers having a very similar number of tokens. If the green line (final distribution) has this characteristic and is significantly more pronounced than the blue line (initial distribution) then the mechanism has done its job in distributing block rewards fairly, and decreasing any overall economic inequality in the population.

Please note when observing distributions in the line chart, that the x-axis is log scale due to the nature of displaying a Pareto distribution on a normal scale, which is far less legible.

\subsubsection{Box Plot}

The box plot provides a summary of the distribution of staked tokens before and after the staking rewards were applied. It displays key statistical measures such as the median and interquartile range (IQR). The chart can be interpreted as follows: the box itself represents the middle 50\% of the data (from Q1 to Q3), with a line inside indicating the median stake amount. The whiskers extend to the smallest and largest values within 1.5 times the IQR, and any outliers beyond this range appear as small circles.

By comparing the initial and final box plots, we can assess how the distribution of tokens changed over the simulated time-frame. The median should always shift upwards in the final distribution, reflecting an overall increase in stake accumulation among stakers. A wider box in the final distribution indicates greater variability in staking rewards, whereas a narrower box suggests that the rewards were distributed more uniformly. The presence of additional outliers in the final distribution could indicate a few stakers receiving significantly larger rewards than others.

This visualization helps in understanding the optimality of token distribution within the SPARC mechanism, i.e., whether rewards are concentrated among a few stakers or more evenly spread across participants. If the final distribution shows a higher median with reduced spread, it suggests stakeholders benefited proportionally. However, if the final distribution exhibits a larger spread with more extreme outliers, it indicates that some stakers received disproportionately higher rewards, possibly due to stake-based advantages.

\subsubsection{Stacked Bar Chart}

The stacked bar chart displays two columns overlaid on top of each other, the initial distribution in blue and the final distribution in green.  The vertical axis represents the difference of average tokens held in lower quartile versus the upper quartile, and both the blue and green bars start at zero on the vertical axis, with whichever bar having the maximum value being positioned behind the other so that both bars are always visible.

The difference between the lower quartile and the upper quartile should be smaller in the final distribution compared to the initial distribution (in terms of the average amount of tokens held). This suggests that the average amount of tokens held increases for smaller validators compared to larger ones.  The intuition is that because we are not considering slashing, and only factoring in block rewards, then the number of tokens for participating stakers can only go up as they accrue rewards.

If the average number of tokens for stakers with the \textit{least amount of stake} to begin with (lower quartile) increases more than the average stake for the stakers with \textit{highest amount of stake} to being with (upper quartile), \textbf{then we are achieving our goal of incentivizing token holders to delegate to the stakers with the lowest amount of stake instead of the stakers with the highest}. This is why we measure the difference between the average stake in the upper quartile versus the lower quartile and compare this difference between the initial and final distributions. If this difference is smaller in the final distribution compared to the initial distribution, it means that the average tokens staked in the lower quartile has increased more than the average tokens staked in the upper quartile.

What we are looking for is the final distribution (Green) to be in front, visible from zero, with the initial distribution (Blue) peeking out from behind over the top of the final distribution (Green). This means that the IQ delta in the final distribution has decreased. In the scenario where the final distribution (Green) is in front and visible from the bottom then the larger the gap between the green and blue bars means the more the average amount of stake for stakers with the least amount of stake to begin with increased compared to those with the largest amount of stake to begin with.

\newpage

\subsection{Discussion of Results}

The following section discussion the results of certain specific simulations, that demonstrate the efficacy of the SPARC mechanism in fairer distribution block rewards in a PoS consensus protocol, when the mechanism is parameterized properly.

\subsubsection{Design Point 0 - Standard PoS - Benchmark}

\begin{figure}[H]
    \centering
    \includegraphics[width=1\linewidth]{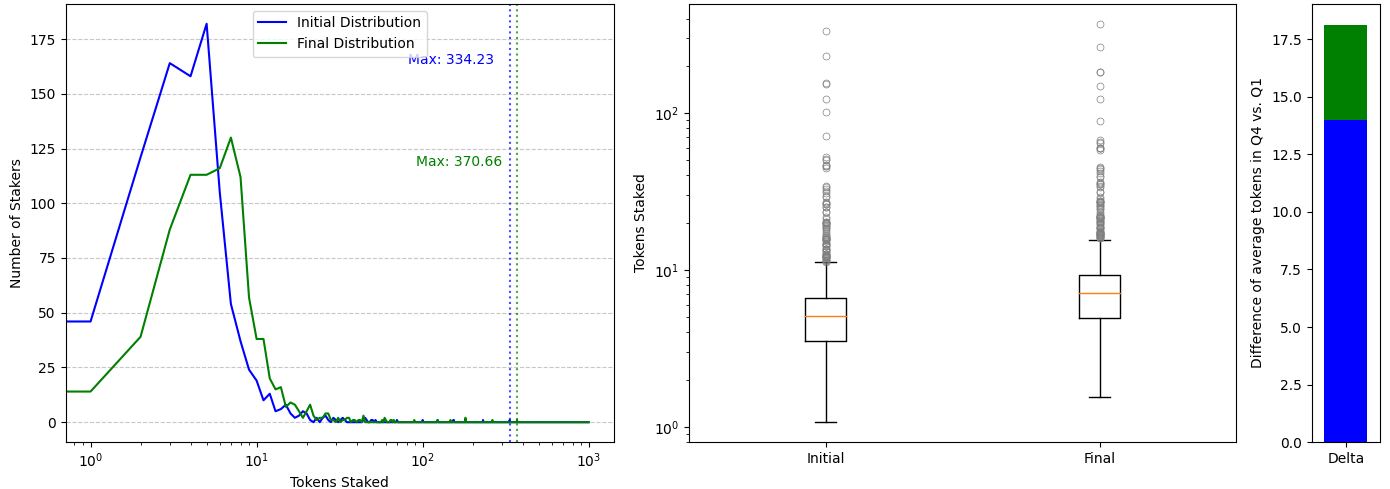}
    \caption{Standard PoS}
    \label{fig:design_point_0}
\end{figure}

We first ran a simulation with a standard PoS protocol in order to benchmark our results against.  Design Point 0 is our benchmarking simulation.  As can be observed from Figure \ref{fig:design_point_0}, the final distribution of tokens across the population is characteristically similar to the initial distribution. These distributions are defined by a large number of stakers having a more than double the amount of other stakers, and with a minority of stakers having almost ten times the amount of the rest of the population.

\subsubsection{Design Point 1}

The following visualizations suggest that a single tier with all 20 stakers receiving the same amount of tokens (i.e., 5\% of block reward each) does not significantly alter the initial distribution. Applying the SPARC mechanism with a single tier maintains the initial distribution almost perfectly.  This is a slight improvement of the vanilla PoS rewards distribution as can be observed from the minimal difference between the blue and green bars in the stacked bar chart, meaning that the difference between the average number of tokens held by stakers in the lower quartile versus the upper quartile was negligible (i.e., the distribution remained the same even the the overall amount of tokens increased).

\begin{figure}[H]
    \centering
    \includegraphics[width=1\linewidth]{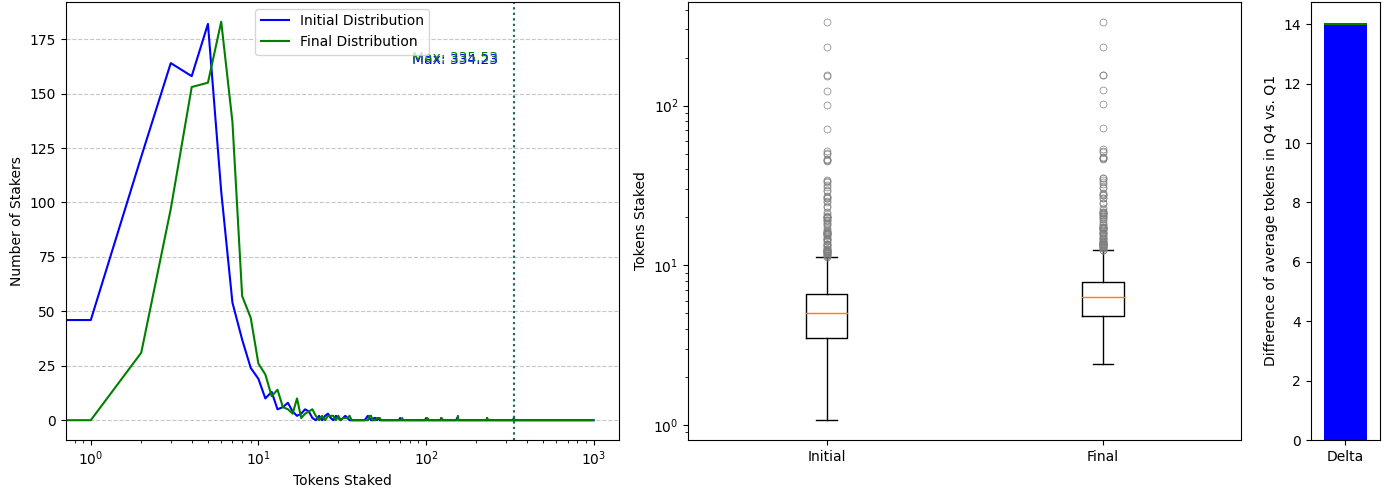}
    \caption{Design Point 1}
    \label{fig:design_point_1}
\end{figure}

\subsubsection{Design Point 2}

This simulation used two tiers, with 10 stakers in each, with a 60:40 split of relative block rewards distributed to each tier. Initially this setup seems to improve the initial distribution slightly, as we can see from the box plot that stakers with lower stake to begin with benefited significantly. The population overall is much better off, however, the shape of the distribution hasn’t changed much. This can be seen by the fact the stacked bar chart shows a negligible difference between the two IQ deltas, with the IQ delta of the final distribution still favoring the stakers with highest stake slightly.

\begin{figure}[H]
    \centering
    \includegraphics[width=1\linewidth]{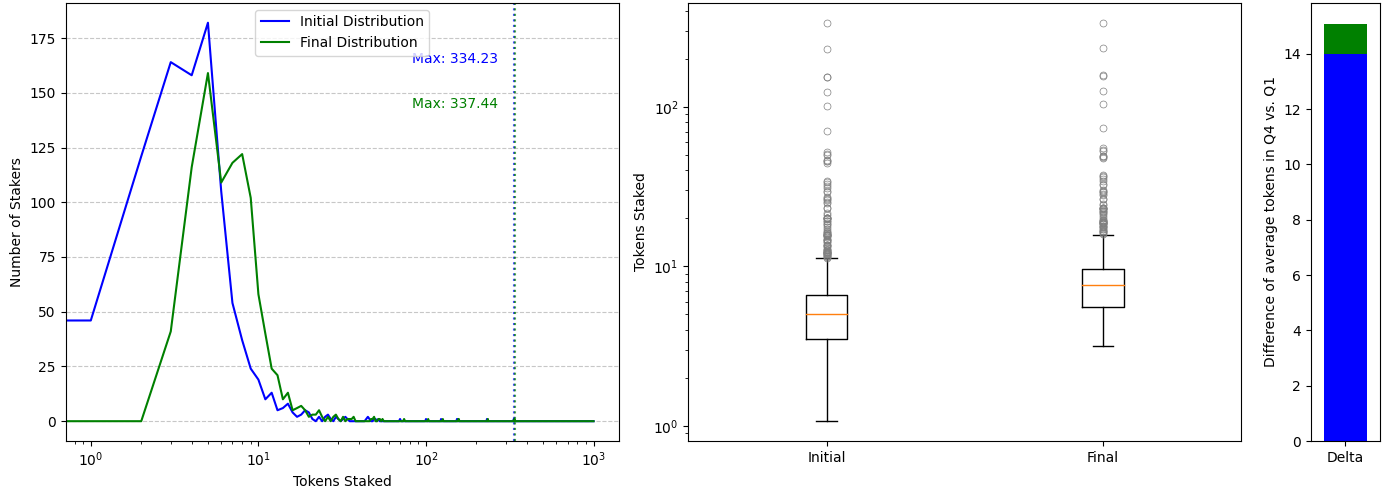}
    \caption{Design Point 2}
    \label{fig:design_point_2}
\end{figure}

\subsubsection{Design Point 6}

This setup improves the distribution slightly insofar as it “squashes” the range of the amount of tokens held and it benefits the stakers that have the lowest initial stake. That being said, it doesn’t show any particularly notable improvement over a standard PoS protocol, except for a an exaggeration of the initial bimodal distribution that can be observed in the two pronounced peaks in the line chart. This setup also uses 3 tiers with 10, 7, 3 stakers in each tier, with 8.3\%, 2\%, 1\% going to each staker in each respective tier.

\begin{figure}[H]
    \centering
    \includegraphics[width=1\linewidth]{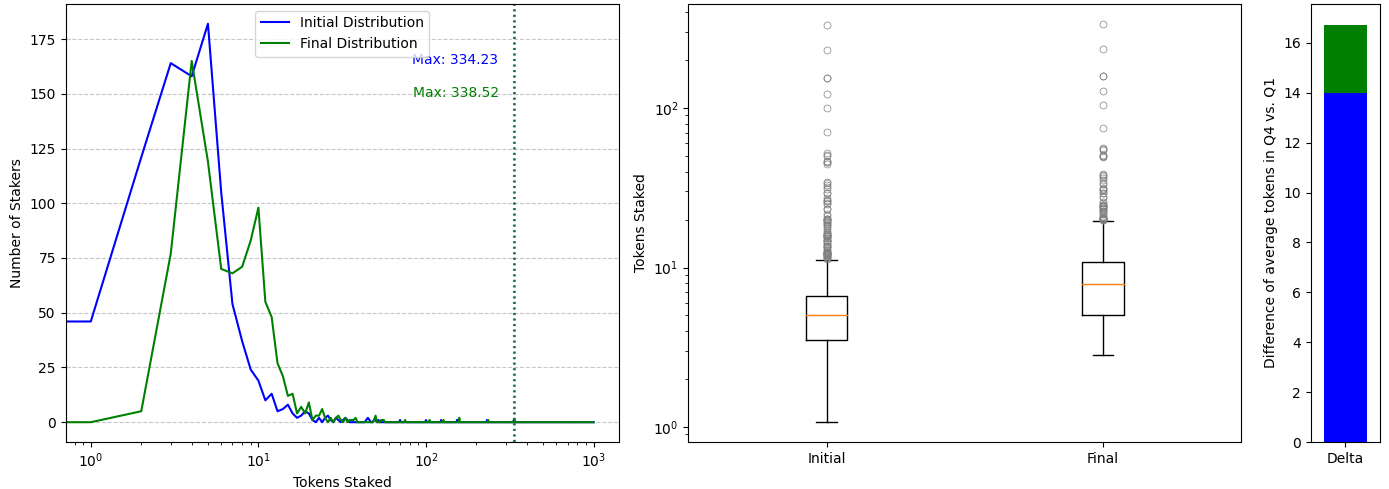}
    \caption{Design Point 6}
    \label{fig:design_point_6}
\end{figure}

\subsubsection{Design Point 9}

The results from this experiment clearly show that the range of the number of tokens held by different stakers has decreased significantly. The final distribution shows that the vast majority of stakers hold more or less the same number of tokens. \textbf{This clearly shows that this specific setup results in more evenly distributed economic power across the population and that over time, stakers are rewarded equally for equal work, regardless of how many tokens they had to begin with}. This can be seen in the single spike in the line chart for the final distribution (green line). The boxplot clearly shows that the inter-quartile range has decreased significantly and the stacked bar chart shows the green line at the bottom / in front (as a reminder: this suggests that the average amount of stake for stakers with the least amount of initial stake \textit{increased} compared to those with the largest amount of initial stake).

This setup was based on 5 tiers with 10, 4, 3, 2, 1 stakers in each respective tier. The share of block rewards to each staker in each tier was 7\%, 4\%, 3\%, 2\%, 1\%.  This suggests that the best results are obtained from putting most amount of stakers in the first tier, and decreasing the number of stakers sub-linearly in each subsequent tier, while at the same time giving most rewards to the stakers in the first tier and and also sub-linearly decreasing rewards to individual stakers in subsequent tiers.

\begin{figure}[H]
    \centering
    \includegraphics[width=1\linewidth]{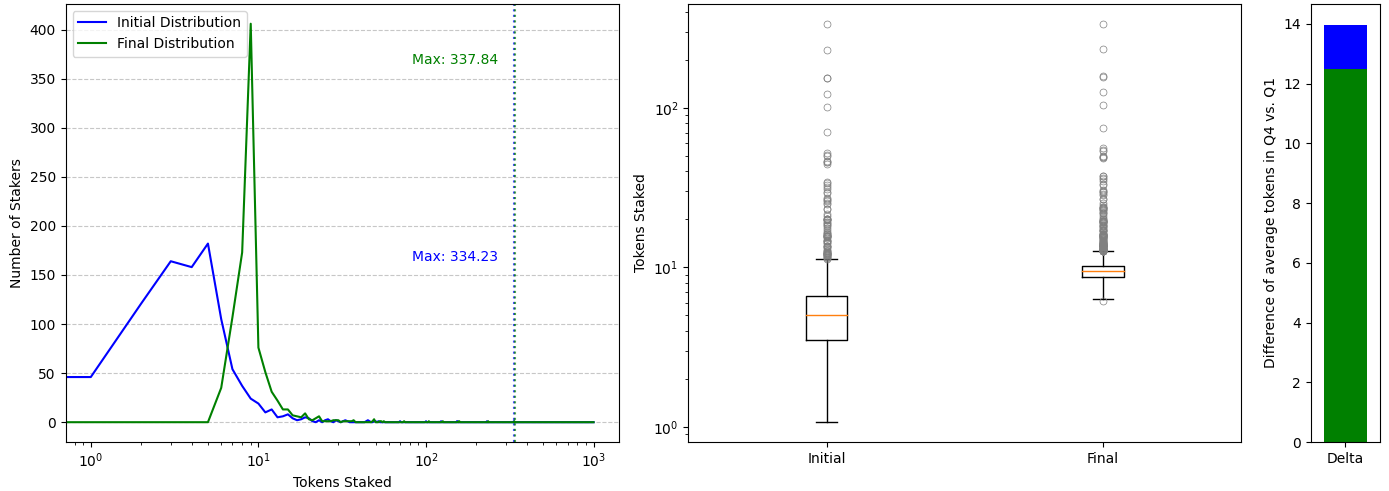}
    \caption{Design Point 9}
    \label{fig:design_point_9}
\end{figure}

\section{Proving that Results are Consistent}

\subsection{Measuring Gini Coefficient across Repeated Simulations}

In order to prove that results of the SPARC mechanism are consistent in improving fairness of distribution of block rewards across the population of stakers, regardless of the initial distribution, we repeated the simulations for each design point and compared the results.

\begin{figure}[H]
    \centering
    \includegraphics[width=1\linewidth]{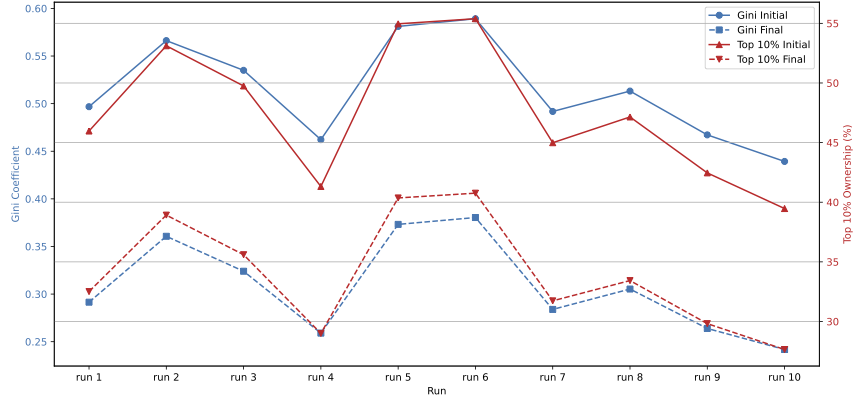}
    \caption{Gini Coefficient vs. Top 10\% Ownership for Design Point 10 - all runs}
    \label{fig:design_point_10_gini_vs_top10}
\end{figure}

The chart in Figure \ref{fig:design_point_10_gini_vs_top10} represents measurements for the initial and final distributions over 10 separate simulation runs for design point 10, which showed very similar results to design point 9, as outlined in Figure \ref{fig:design_point_9}. Each repeated run is plotted on the x-axis, and for each run we measure both the gini coefficient (on the left axis) and the top 10\% of ownership of tokens (on the right axis).

As can see observed in Figure \ref{fig:design_point_10_gini_vs_top10}, the results are consistent throughout, even with a randomized initial distribution. In each case the gini distribution is lowered significantly, from what would normally be classified as highly concentrated economic power, to much more evenly distributed.

The percentage of stakers that own the top 10\% of tokens decreased in each run, a result which considered in isolation would suggest an increase in inequality, with fewer stakers owning the top 10\% of the token supply.  However, what this is actually representing is a decrease in the number of outliers in the distribution, with many of the stakers that were initially in the top 10\% of token holders instead having a similar number of tokens to other stakers in the final distribution.

\subsection{Mean and Median Tokens Staked across Repeated Simulations}

\begin{figure}[H]
    \centering
    \includegraphics[width=1\linewidth]{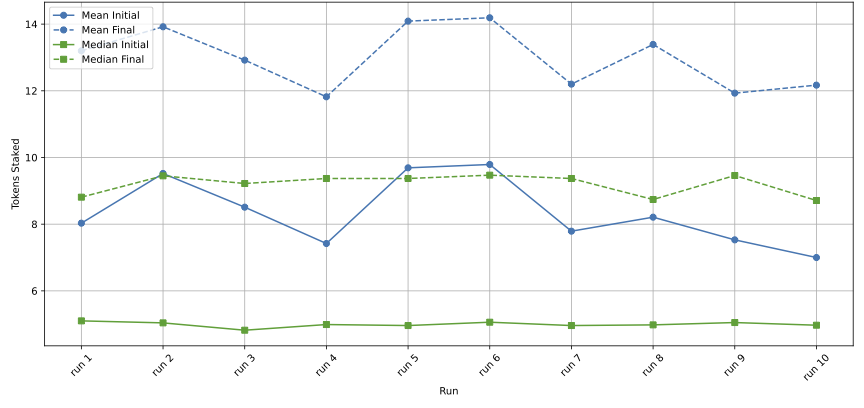}
    \caption{Mean and Median Tokens Staked for Design Point 10 - all runs}
    \label{fig:design_point_10_mean_median}
\end{figure}

The chart in Figure \ref{fig:design_point_10_mean_median} shows both the mean and median amount of tokens for both the initial and final distribution of tokens for design point 10 across the same 10 repeated simulations as visualized in Figure \ref{fig:design_point_10_gini_vs_top10}.

As can be observed from the visualization, both the mean and median are significantly reduced in the final distribution in each case. This is despite the fact that there are no slashing penalties implemented and therefore the overall number of tokens increases in the final distribution. This again implies that tokens are more evenly distributed across the population, without having to re-distribute tokens from any individual validator to another.

\subsection{Gini Coefficient across all Design Points and Simulation Runs}

\begin{figure}[H]
    \centering
    \includegraphics[width=1\linewidth]{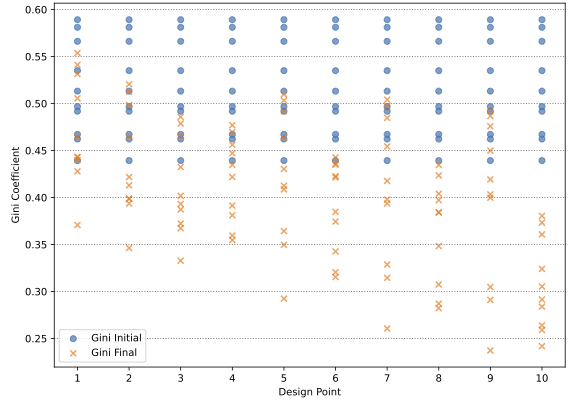}
    \caption{Gini Coefficient across all design points and runs}
    \label{fig:gini_scatter_all_design_points}
\end{figure}

The scatterplot chart in Figure \ref{fig:gini_scatter_all_design_points} shows every design point and results from 10 repeated simulations for each. In each simulation, there was a randomized initial distribution, represented by the blue dots, with the final distribution marked as orange crosses.

As can be observed from the visualization, there are varying results between design points. Certain experiments deliver much better results then others, most notably design points 7, 9 and 10. This indicates that the mechanism is quite sensitive to the specific parameterization, and that care must be taken when designing the protocol. The best results appear to come from design points 9 and 10, which have similar setups. Both have 5 tiers with a descending order of rewards per validator per tier.

\section{Discussion and Future Work}

\subsection{Alternate to Tiered Approach}

An alternative to using the tiered approach to block rewards distribution is to generalize the distribution of rewards as a decreasing function over validator rank using an inverse power decay formula. Each validator is ranked by their stake (i.e., highest stake gets rank 1), and the reward weight for the $i^{\text{th}}$ validator is:

\[
r(i) = \frac{1}{i^p}
\]

To normalize these weights so that the total reward distributed is $R$, we compute the normalized reward for each validator as:

\[
R_i = R \cdot \frac{r(i)}{\sum_{j=1}^{n} r(j)}
\]

Where:

\begin{itemize}
    \item $p > 0$ is a tunable exponent controlling the skew of the distribution (e.g., $p = 1.5$).
    \item $n$ is the size of the validator committee (e.g., $n = 20$).
    \item $R$ is the total reward to be distributed among the committee for the slot.
\end{itemize}

This formulation produces a smooth, continuous, rank-sensitive reward curve that avoids the rigidity and complexity of tier-based systems. It encourages higher staking without requiring arbitrary thresholds.

The benefits of this alternate approach are that it avoids the complexity of tier definitions, replacing it with a simple and easily tunable decay function.  This also has the benefit of encouraging coopetition across the whole committee rather than only within tiers.  However, there is a trade-off: the decay function may be less intuitive to stakeholders, and the differentiation among lower-ranked validators can potentially become negligible, which could weaken incentives for marginal stake improvement. Ultimately, the trade-off can be thought of as being between explicit control and transparency (tiered) and algorithmic simplicity and continuity (decay-based).

\subsection{Anti-Sybil Considerations}

Like many cryptoeconomic mechanisms, this design is susceptible to Sybil attack \cite{leshno2021prior}.

A well-capitalized rational entity will be incentivized to maximize their rewards by spreading their staked tokens across as many new validators as possible, as the reward mechanism is designed to decentralize economic distribution by providing outsized rewards for user nodes with minimal stake.

This only applies to validators that are delegated stake from third party delegators, as they will stake with validators that have lower total stake, seeking to avoid reward dilution. In this respect, a single entity may gain additional delegation by splitting their stake across multiple validators that they control. In such cases, multiple Sybil validators may collectively accrue rewards comparable to those of a single, high-stake validator.

We can model Sybil resistance in an instantiation of SPARC by comparing the expected reward of a validator with stake \( S \) versus that of \( m \) Sybil instances each holding \( S/m \). Let \( p_j(S) \) denote the probability that a validator of stake \( S \) is placed in tier \( j \), with per-tier reward \( R_j \) and tier size \( x_j \). The expected reward of a single validator is:

\[
\mathbb{E}[R(S)] = \sum_j p_j(S) \cdot \frac{R_j}{x_j}
\]

For \( m \) Sybil validators, each with stake \( S/m \), the expected cumulative reward becomes:

\[
\mathbb{E}[R_{\text{Sybil}}] = \sum_{k=1}^m \sum_j p_j(S/m) \cdot \frac{R_j}{x_j}
\]

If \( \mathbb{E}[R_{\text{Sybil}}] > \mathbb{E}[R(S)] \), Sybil behavior is rational; otherwise, it is disincentivized.

The design of SPARC can be tuned to maintain Sybil resistance by configuring narrow, steeply rewarded tiers, reducing the number of validators per top tier, or incorporating diminishing returns mechanisms. However, these properties would need to be robustly simulated and tested when applying SPARC mechanism to a PoS protocol.

\subsubsection{Potential Sybil Resistance Augmentations}

In order to prevent this system from being abused, it is recommended that a process for network operator entry be deployed which minimizes the probability of successful Sybil behavior. This could be accomplished by the network introducing a governance process to maintain the  efficacy of SPARC's continuous economic decentralization mechanism which motivates the rewards system. For example, a DAO could periodically elect members to an 'Anti-Sybil Council' to ensure independence of each operator from one another, and thereby the DAO will gate network participation by proxy. A direct plutocratic vote on each operator's entrance might not be tenable, as the council's job would be to vet prospective operators while also ensuring participants are not publicly doxxed as would be the case if the DAO directly gated admission.

There is also a potential for decentralized ID (DID) and/or zk-based solutions \cite{sánchez2024zeroknowledgeproofofidentitysybilresistantanonymous} to be brought to bear to validate entity uniqueness while preserving privacy, especially when coupled with privacy-preserving biometrics, for example as proposed by Humanode \cite{kavazi2021humanode}. There have also been proposed various other approaches to strengthen Sybil resistance in PoS networks \cite{BIRYUKOV2020101109} \cite{PLATT2021108424} \cite{10140278} which could potentially be employed. Although this implies that node operation is not permissionless in the traditional sense, as long as an operator is not attempting to Sybil attack the network, then they should be able to participate.

\newpage

\section{Conclusion}

The SPARC mechanism introduces a novel approach to reward allocation in proof-of-stake consensus by combining randomized committee selection with deterministic, stake-ranked tier assignments. This design enables fine-grained control over incentive gradients while preserving the fair and optimal distribution of rewards over time. By departing from proportional stake-based rewards and instead rewarding validators based on intra-committee performance, SPARC mitigates centralization pressures and creates space for diverse participation across the validator set.

The SPARC mechanism offers a compelling approach to enhancing decentralization and economic security in proof-of-stake systems. By incentivizing stake delegation and introducing gamification, it addresses key challenges and fosters a fairer distribution of economic power, reducing centralization risks. Future research will focus on refining the emission curve, reward structures, and Sybil resistance strategies to optimize the system's performance and resilience.

\printbibliography

\end{document}